# Collimating Slicer for Optical Integral Field Spectroscopy

Florence Laurent*[a], François Hénault[b]

[a]Univ Lyon, Univ Lyon1, Ens de Lyon, CNRS, Centre de Recherche Astrophysique de Lyon UMR5574, F-69230, Saint-Genis-Laval, France;
[b]Institut de Planétologie et d'Astrophysique de Grenoble, Université Grenoble-Alpes, Centre National de la Recherche Scientifique, B.P. 53, 38041 Grenoble, France

**ABSTRACT**

Integral Field Spectroscopy (IFS) is a technique that gives simultaneously the spectrum of each spatial sampling element in a given object field. It is a powerful tool which rearranges the data cube (x, y, λ) represented by two spatial dimensions defining the field and the spectral decomposition in a detector plane. In IFS, the "spatial" unit reorganizes the field and the "spectral" unit is being composed of a classical spectrograph. The development of a Collimating Slicer aims at proposing a new type of integral field spectrograph which should be more compact. The main idea is to combine the image slicer with the collimator of the spectrograph, thus mixing the spatial and spectral units. The traditional combination of slicer, pupil and slit elements and the spectrograph collimator is replaced by a new one composed of a slicer and collimator only. In this paper, the state of the art of integral field spectroscopy using image slicers is described. The new system based onto the development of a Collimating Slicer for optical integral field spectroscopy is depicted. First system analysis results and future improvements are discussed. It finally turns out that this new system looks very promising for low resolution spectroscopy.

## 1. INTRODUCTION

Integral Field Spectroscopy (IFS) is a technique that gives simultaneously the spectrum of each spatial sampling element in a given Field of View (FoV). It is a powerful tool which rearranges the data cube (x, y, λ) represented by two spatial dimensions defining the field and the spectral decomposition in a detector plane (X", Y"). In IFS, the "spatial" unit reorganizes the field while the "spectral" unit is being composed of a classical spectrograph. The spatial unit acts as a link between the telescope and the spectrograph by reformatting the FoV. Three main techniques – microlens array, microlens array associated with optical fibres and image slicer – are used in astronomical instrumentations (Figure 1). The advantage of image slicer system compared to the others is data-packing efficiency which is around 90% and which is also more compact. The principle of an image slicer system is based on the concept proposed by R. Content in 1997 [1] and illustrated in Figure 4, top panel. The two main optical functions of the image slicer are to:
- Transform a rectangular FoV in plane OXY into a series of mini-slits located at the entrance focal plane of the spectrograph O'X'Y',
- Reimage the telescope pupil (usually located at infinite distance) at the spectrograph pupil P'U'V'.

Therefore, the light from each mini-slit is dispersed to form spectra on the detector and a spectrum can be obtained simultaneously for each spatial sample within the FoV. More precisely, an image slicer is originally composed of three reflective elements: a slicer stack, rows of pupil and slit mirrors.

- The slicer stack (in the OXY plane). It is composed of a stack of several thin spherical or flat mirrors (called "slices") slicing the field and reflecting real images of the telescope pupil on the pupil mirrors.
- The pupil mirrors are disposed along a row parallel to the spatial direction (the U-axis in Figure 4, top panel). Each pupil mirror then re-images its associated slice on the corresponding slit mirror located at the spectrograph's focal plane (slit plane O'X'Y').
- The slit mirrors are also disposed along a row parallel to the spatial direction (the X'-axis). Finally, each slit mirror acts as field lens, re-imaging the telescope pupil (which is onto pupil mirrors) onto the entrance pupil of the spectrograph.

*florence.laurent@univ-lyon1.fr; phone +33 4 78 86 85 33; fax +33 4 78 86 83 86; http://cral.univ-lyon1.fr/

The subject of this paper is to describe a modified image slicer linked with the spectrograph collimator. Section 2 firstly introduces the state of art of IFS in past and future astronomical instrumentation. Section 3 summarizes different existing optical designs for image slicer. A new proposal for Integral Field Spectroscopy called "Collimating Slicer" is described in section 4. Finally, section 4.3.2 compiles the conclusions and future developments.

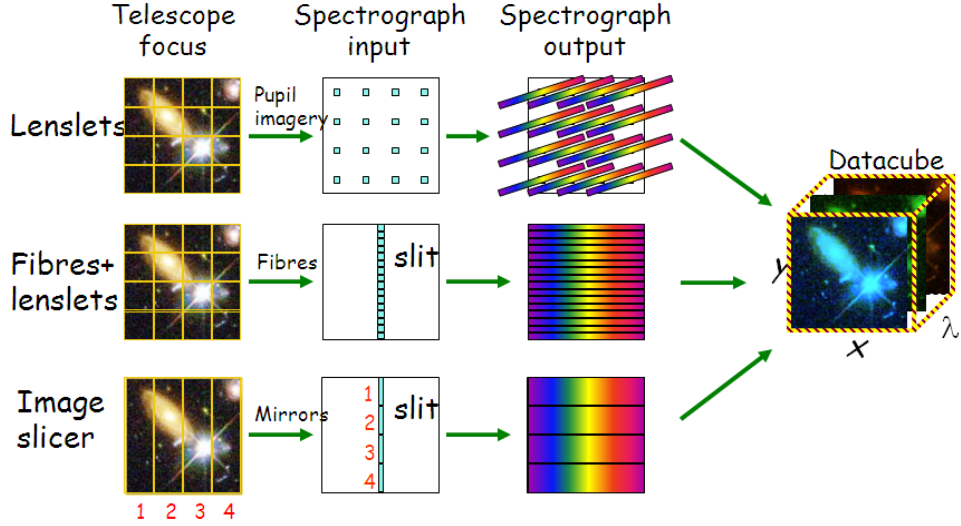

Figure 1: Principle of Integral Field Spectroscopy using three techniques – microlens array, microlens array associated with fibers and image slice [1].

## 2. STATE OF ART FOR INTEGRAL FIELD SPECTROSCOPY WITH IMAGE SLICER

As seen in the previous section, image slicers are made of different mirrors with different geometrical characteristics (tilts and curvature radius). Moreover, different materials, manufacturing and assembling processes can be employed, including glass contacting or metal diamond turning. Throughout the years, progresses in those technologies came along with larger telescope diameters offered by the observatories, and the availability of bigger detector arrays composed of thousands of rows and columns in different spectral bands of interest. The goal of this state of art is to give an overview of the main existing and future IFS using the image slicer concept. For that purpose different parameters are selected:

- Instrument Name and its associated telescope with telescope diameter ($D_{Tel}$)
- Field of View ($\theta$)
- Wavelength range ($\Delta\lambda/\lambda$)
- Sampling onto the sky ($d\theta$)
- Number of spatial pixel ($N_\theta$)
- Number of spectral pixel ($N_R$)
- Spectral resolution (R)
- Geometrical *étendue* (G=A.$\theta$ where A is the collecting area)
- Spectral Power ($P_S$=R.$N_R$)

After collecting these data the main existing and future IFS using image slicers could be classified following several criteria as presented in Figure 2. Due to the increase of CCD size, we note that the number of spatial and spectral pixels increases for future projects. Moreover, the spectral resolution is between 300 and 22000 for a sampling lower than 1 arcsec². For the future projects as Harmoni [2] or Frida [3], the sampling decreases. Due to the increase of telescope diameter, the spectral power increases significantly.

Another conclusion drawn from the charts in Figure 2 is the fact that it exists very few instruments operating simultaneously at a relatively high angular resolution and a rather modest spectral resolution, typically lower than 100.

This particular domain of investigation is indicated by dashed boxes in the Figure 2. Before exploring it in section 4, we present a brief overview of image slicer designs in the next section.

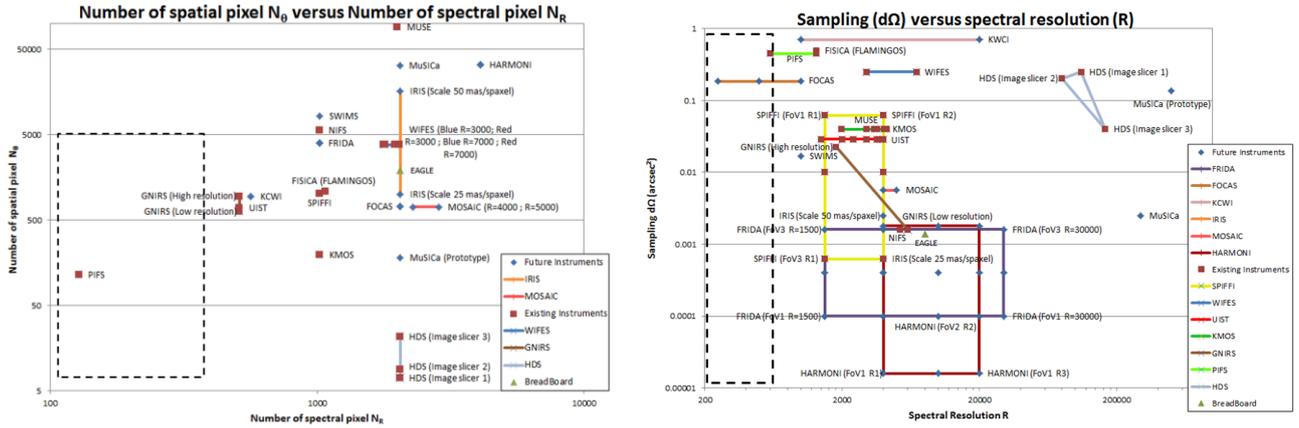

Figure 2 : Graphs for existing and future instruments with image slicer concept. Left: Graph presenting the numbers of spatial versus spectral pixels. Right: Graph presenting the sampling versus spectral resolution.

## 3. A BRIEF HISTORY OF IMAGE SLICING

### 3.1 Pioneering works

In 1938, I. S. Bowen was the first to imagine the image slicer concept [4]. It consisted in a stack of flat mirrors coupled with a cylindrical lens and creating a pseudo-slit at the spectrograph entrance. The resulting device allowed decreasing the width of the spectra. Bowen's image slicer was manufactured by Pierce in 1965 [5]. Sixty years after, Walraven proposed a modified version of this slicer where the mirrors are replaced by total reflections inside parallel plates which are glued by optical contact on a prism. This concept was manufactured by Simmons in 1980 [6]. In 1980, Richardson improves the previous concept by introducing multiple reflections between a pair of concave mirrors [7]. Nevertheless, the purpose of all these image slicers was not integral field spectroscopy in astronomy. It is not before 1996 that the concept was applied to IFS with the development of the 3D instrument equipped with flat image slicers.

### 3.2 Flat Image Slicer

The 3D near infrared imaging spectrometer [8] represented a new generation of astronomical instrumentation. It was based on a 256² NICMOS-3 Rockwell array and could simultaneously obtain 256 H-or K-band spectra at R= 1100 or 2100 from a square field of 16x16 pixels on the sky. The main specifications of the instrument are indicated in Table 1, together with those of the other instruments discussed in this section.

Table 1: Sample specifications of typical IFS instruments equipped with image slicers.

| SPECIFICATIONS | INSTRUMENT | | | | Unit |
| --- | --- | --- | --- | --- | --- |
| | 3D | IFMOS | MUSE | MuSICa | |
| Field of view (FoV) | 4.8 x 4.8 | 40 x 40 | 60 x 60 | 9 x 9 | arcsec |
| FoV sample | 0.3 / 0.5 | 0.2 | 0.2 | 0.05 | arcsec |
| Spectral band | H and K bands | 1 - 5 | 0.465 - 0.93 | 3.9 - 25 | microns |
| Spectral resolution | 1100/2100 | 1000 | 2000 - 4000 | 300000 | |
| Slices number | 16 | 2 x 30 | 24 x 36 | 8 x 16 | |
| Slice width | 0.4 | 0.9 | 1.6 / 1 | 0.05 | mm |
| Pixel number | 256 x 256 | 2048 x 2048 | 4096 x 4096 | 4096 x 4096 | |
| Pixel size | 16 | 18 | 15 | 10 | microns |

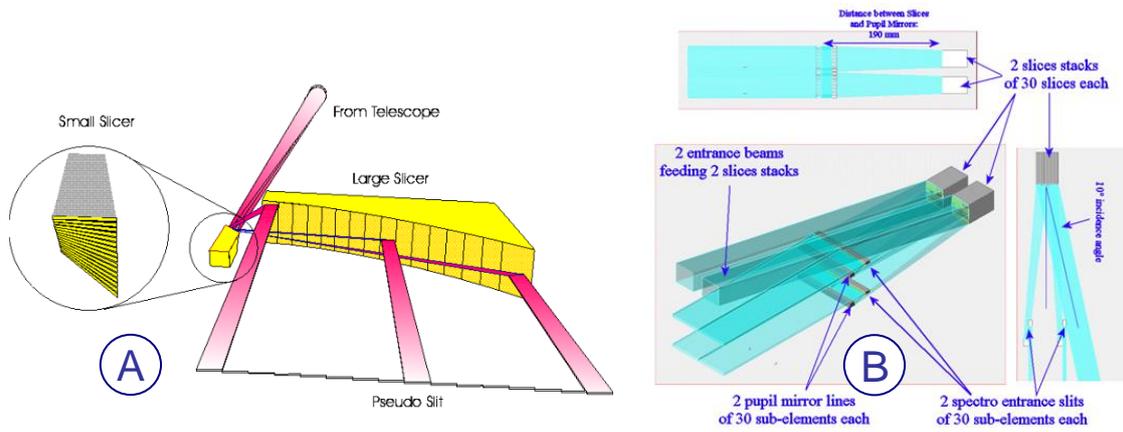
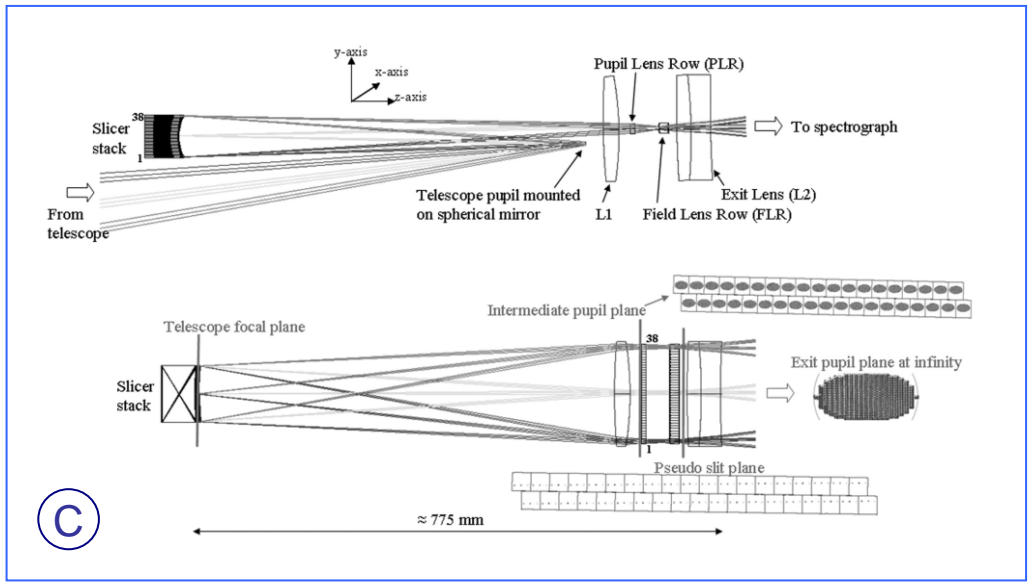
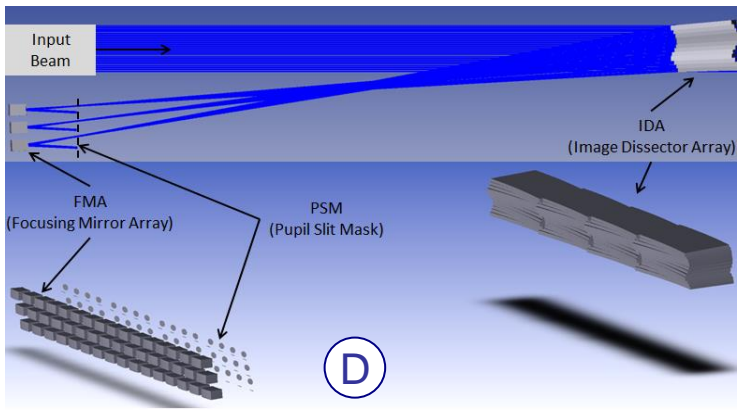
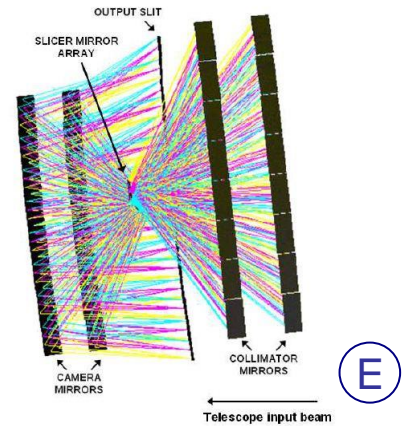

Figure 3 : Illustrating various concepts of image slicers. (A) Flat image slicer of the 3D instrument. (B) ESA IFMOS instrument. (C) General optical layout of the Catadioptric Image Slicer. (D) MUSE image slicer (E) Multi-Slit Image Slicer of MuSICa.

The Flat Image Slicer uses a system of plane mirrors to transform the two-dimensional FoV of the instrument into a pseudo-slit. It consists of two sets of mirrors (Figure 3-A). The first one, the small slicer, is a stack of 16 flat mirrors placed in the focal plane of the telescope. The mirrors are only 0.4 mm thick and each mirror is tilted by a different angle around the same direction (the Y-axis of Figure 4, top panel). The 16 different tilt-angles lead to a sliced field-of-view consisting of 16 stripes fanned out over >90°. The projected length of the 16 stripes in the telescope focal plane is 6.4 mm, corresponding to 16 pixels of 0.4 mm width. After the small slicer, the fan of 16 stripes hits the second set of mirrors, called the large slicer, that reflects the rays to form a pseudo-slit. The tilt-angles of the 16 mirrors of the small and large slicer are such that the position of the telescope pupil is conserved. To achieve this, the centres of the large slicer mirrors have to lie on a hyperbola. The large slicer mirrors are placed at a distance from the small slicer where the 16 stripes do not overlap anymore, resulting in a contiguous pseudo-slit of 16 slitlets arranged in a staircase pattern.

### 3.3 R. Content's Image Slicer

Despite the successful achievement of the 3D instrument, the concept of flat image slicer remains limited by its exclusive use of flat mirrors, inducing excessive length of the pseudo-slit and geometrical constraints related to pupil locations. The original concept of Advanced Image Slicer proposed by R. Content in 1998 allows removing these drawbacks by replacing the flat mirrors with concave mirrors. Figure 4, top panel gives a general illustration of this design that was subject to several adaptations in order to fulfill various spectroscopic requirements. Details of the operational principle may be found in [1]. Here is given an example of the main features of the design principle and those details which are directly relevant to a prototype proposed to ESA for the near-infrared spectrograph NIRSpec of the future James Webb Space Telescope (JWST). The baseline for the optical design of the ESA prototype was the design of the IFMOS instrument. This latter had a 40" x 40" field of view sampled at 0.2" (see the main optical requirements in Table 1). To achieve this, the IFMOS included four identical spectrographs, each using a pair of 30 slicing-mirror stacks to slice and rearrange the input FoV into two exit slits at the entrance of the spectrograph. These stacks were followed by lines of pupil and slit mirrors as depicted in Figure 3-B. The ESA prototype has been developed by a consortium led by LAM and including the CRAL and the Durham University since 2002. Successful cryogenic tests were performed in 2004, demonstrating the maturity of the concept for space applications [9].

### 3.4 Catadioptric Image Slicer (CIS)

In the Catadioptric Image Slicer, pupil and slit mirrors are replaced by dioptric elements. This allows to improve both image performances and costs. Nevertheless, dioptric elements present chromatic aberrations and result in a complex arrangement of pupil and slit elements since they must be arranged along two adjacent rows. Moreover, it requires additional field lenses. This concept was originally proposed for the MUSE instrument [10][11][12]. MUSE is a second-generation integral-field spectrometer which has been proposed to the European Southern Observatory (ESO) for the Very Large Telescope (VLT). It combines a 1'x1' field of view with the spatial resolution of 0.2 arc seconds and it operates in the visible and near infrared (see Table 1). The FoV is divided to 24 sub-fields by means of a field-splitter and each sub-field is sent to an image slicer combined with a spectrograph. The CIS was composed of 36 mirror slices, two rows of mini lenses (the pupil lens and field lens rows in Figure 3-C) and two field lenses (L1 and L2). A prototype has been manufactured in 2003-2004 and successful performance tests were carried out in 2004.

### 3.5 MUSE Image Slicer

In the final MUSE image slicer, the pupil and slit elements were combined into a single one, therefore the system only comprises two major optical components: the slicer stack and a row of focusing mirrors [13]. The MUSE slicer reorganizes the subfield of view of 2.5x60 arcsec² into a pseudo-slit of 0.2 arcsec width. The image slicer is composed of one Image Dissector Array (IDA) which splits the subfield in 48 slices (4 identical stacks of 12 slices) and creates 48 intermediate pupils of the VLT following a "staircase" arrangement. Each intermediate pupil is sent to a focusing mirror located near the pupil plane. The so-called Focusing Mirror Array (FMA) forms an image of each individual slice in the pseudo-slit plane O'X'Y' and re-images all the intermediate telescope pupils at the entrance pupil of the MUSE spectrographs. The exit slit plane is arranged in three staggered rows of 48 mini-slits (see Figure 3-D). A Pupil/Slit mask (PSM) between the IDA and FMA prevents possible ghosts and stray light. It is constituted of 48 elliptical apertures located at the intermediate telescope pupils and of 48 rectangular holes in the O'X'Y' plane. The 24 MUSE image slicers were manufactured following this optical design.

### 3.6 Multi-Slit Image Slicer

The Multi-Slit Image Slicer is an interesting variant of the Flat Image Slicer: it makes use of plane slicing mirrors associated with two arrays of spherical mirrors (one collimator mirror array and one camera mirror array) to transform the two-dimensional FoV into a pseudo-slit. It will be implemented on MuSICa (Multi-Slit Image slicer based on Collimator-Camera), a solar integral field spectrograph for the 4-m aperture European Solar Telescope [14]. MuSICa is a multi-slit image slicer that decomposes an 80 arcsec² field of view into slices of 50 µm and reorganizes it into eight slits of 0.05 arcsec width and 200 arcsec long. This is a telecentric system with an optical quality at diffraction limit compatible with different modes of operation of the spectrograph.

Since the slicer mirrors of MuSICa are flat, the intermediate pupil images are formed by the collimator mirrors that are spherical. Because of the antisymmetric correspondence of collimator and camera mirrors, the intermediate pupil images are made to overlap. The output slit is generated in the middle of the two camera mirrors columns alternating focusing beams of each column to compensate the angles. The whole optical design is presented in Figure 3-E.

### 3.7 Conclusion

Several adaptations have been developed in order to fulfil various requirements: flat or spherical mirrors, one or two arrays of secondary mirrors, combination of mirror and lens, various materials, different manufacturing and assembling processes. In all cases, the image slicer concept presents important advantages when compared to fiber or lenslet coupling. The main advantages are:
- By allowing each of the optical surfaces to have power, the Image Slicer brings important benefits in terms of compactness,
- By preserving all spatial information,
- By working in cryogenic instruments and at long wavelengths.

In the following is described the Collimating Slicer design, which is particularly well suited to high angular resolution and moderate spectral resolution IFS. It is obviously inspired by all the preceding designs, and more specifically by the use of dioptric elements as for the Catadioptric Image Slicer, and of a multiple lens camera array as in the Multi-Slit Image Slicer.

## 4. COLLIMATING SLICER

### 4.1 Description

The classical approach for giving simultaneously the spectrum of each spatial sampling element of a given field in a detector plane is illustrated by the upper panel of Figure 4. It simply consists in using two sub-assemblies:
- An image slicer which transforms a rectangular FoV in a series of mini-slits located at the entrance focal plane of the spectrograph and reimages the telescope pupil at or near infinite distance onto the spectrograph pupil.
- A long-slit spectrograph composed of a collimator, a diffraction grating, a camera and a detector array.

The IFS with image slicer is typically composed of seven major components where the pupil is "moved" three times:
- First, by the slicer stack. Each slice reflects the incident rays in a different direction of the XZ plane, toward its associated pupil mirror, where an intermediate image of the telescope pupil is formed.
- Second, by the slit mirrors. Their main optical function is to re-image all the intermediate pupils coming from pupil mirrors at the entrance pupil of the Spectrograph (i.e. at infinity), thus allowing a significant reduction in the diameters of its optics. In fact these slit mirrors act as field lenses in classical optical systems.
- Third, by the spectrograph collimator. The collimator images the common pupil onto the diffraction grating.

However, this optical layout suffers from several drawbacks:
- The length of the system can be excessive (typically larger than one meter), compromising its using in future IFS for large telescopes where several paths are mandatory,
- There is a large number of optical components. The image slicer is composed of 3 arrays of N mirrors depending on spatial and spectral sampling. For example, as presented onto the upper panel of Figure 4, the image slicer is composed of 4x3 optical components. In MUSE instrument, 24x48x2 elements were manufactured. In addition, their manufacturing and assembling must be very accurate.

- Due to the necessity to have a common pupil onto the diffraction grating giving high spectral resolution, the size of the optics are generally larger than 250mm. A custom made manufacturing is required.

In opposition, the main idea of the collimating slicer is to "merge" the spectrograph collimator with the pupil and slit mirrors in order to image the pupil only once rather than three times. The proposed system is composed of the slicer mirror and of a collimating lens. Then, the IFS can be more compact (typically shorter than 0.3 meter) and with less optical components. No common pupil has to be reimaged allowing smaller camera lenses with standard optical elements. The latter are schematically illustrated in the lower part of Figure 4, indicating the main coordinate frames. Following the sense of the incoming light, they are successively:

- The O'X'Y' plane of the slicer stack. It is placed at the image plane of the telescope (or at any other conjugated plane within the optical system). The stack is constituted of N very thin spherical or flat mirrors called "slices", whose lengths are parallel to the X'-axis and thin dimensions are parallel to the Y'-axis. The slices are stacked together along the Y'-axis, then constituting a monolithic block whose rectangular contours correspond to the dimensions of the entrance FoV. Ideally there should be no "dead areas" between each slice, so that no significant region of the observed FoV is lost. Each slice has different orientations around the X' and Y' axes in order to reimage the intermediate pupils at different locations arranged as an array.

- The $P_-'U_-'V_-'$ plane of the Collimating Lens. This lens is essential for collimating the FoV onto the diffraction grating. To respect the condition of telecentric field onto the diffraction grating, the distance between the slicer stack and collimating lens is equal to its focal length. The choice of its focal length should be optimal and is very critical because it is directly linked to:
    - The pupil diameter onto the diffraction grating which allows to control the encumbrance of the system,
    - The focal length of the camera and required spatial sampling,
    - The spectral resolution and diffraction angle.

- The P'U'V' plane of the diffraction grating. Each intermediate pupil is imaged onto the diffraction grating. The distance between the telecentric collimating lens and diffraction grating corresponds to focal length of collimating lens, because the slices are flat.

- The $P_+'U_+'V_+'$ plane of the microlens camera. Together with the collimating lens, it forms the image of the slicer stack onto the detector array. This camera is actually composed of N microlenses corresponding to the same number of slices with the same orientation in array. The distance from the diffraction grating should be small due to diffraction angle. The focal length of the microlenses is linked to the collimating lens and the required spatial sampling. In this case, the focal length should be quite small (lower than 20mm).

- The O"X"Y" plane of the detector array, where are formed N images of the N slices arranged in array, each with its corresponding spectrum.

This theoretical description can be considered as the starting design rules for a collimating slicer. Stringent requirements are set on the pupil arrangement onto the diffraction grating in order to optimize spectra positioning onto the detector. Moreover, the choice of the collimating lens focal length is also very important because it determines the global encumbrance of the system. In order to assess the feasibility of this type of collimating slicer, we decided to realize a real case study, whose main requirements are defined in the following sub-section.

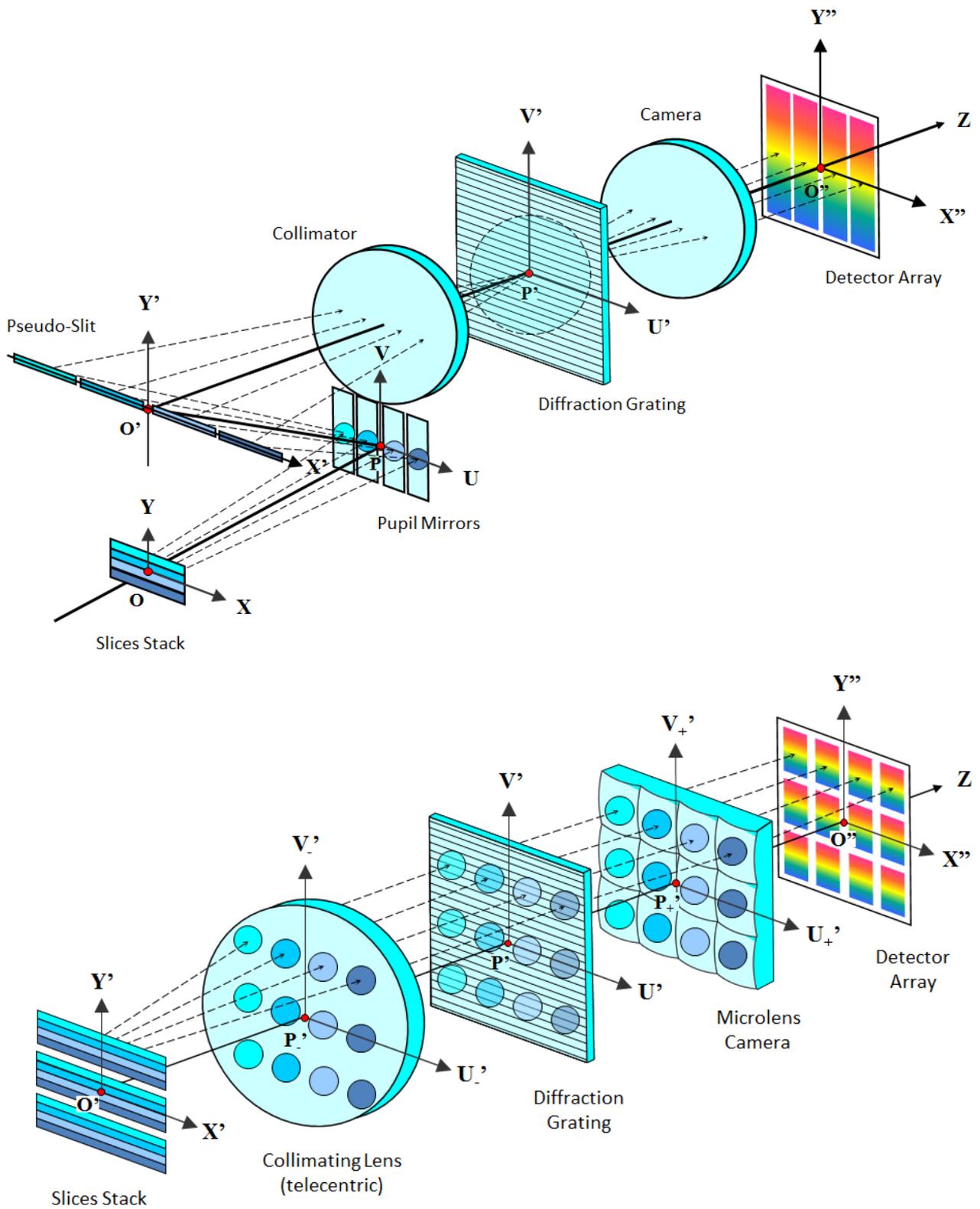

Figure 4 : Schematic views of Integral Field Spectroscopy: with classical image slicer (top panel) and with the proposed Collimating Slicer (bottom panel).

## 4.2 Requirement for a case study

In order to demonstrate the opportunity of an integral field spectroscopy based on collimating slicer, we worked on an example coming from possible real-world astronomic case. It could be used in extragalactic domain, to observe Ly-α line and to measure galaxy redshift with the continuum. Here the basic idea is to convince the reader of the feasibility of collimating slicer, opening the way to use standard components in IFS and to reduce the mechanical encumbrance of classical spectrographs.

In order to keep more reasonable geometrical parameters, it was then chosen to:
- Start with a 4m class telescope diameter as the Calar Alto telescope (3.5m diameter, f#=3.93)
- Limit FoV at 0.2x0.07 arcmin² with spatial sampling of 0.2arcsec²
- Introduce an anamorphic ratio along y-axis to respect the Nyquist criteria onto the detector
- Take wavelength range as bluer as possible in order to cope with small spectral resolution scientific case. The wavelength range is 350-750nm
- Limit spectral resolution below 500 in order to limit microlens camera size
- Fix the slice thickness to 0.5mm which is considered as a limit for glass manufacturing
- Fix the collimating focal length to 100mm to have an intermediate pupil diameter with the same order of dimension as the mini slit (intermediate pupil diameter of ~1mm)
- Chose an arrangement in 2 rows of 10 images onto the detector
- Use a standard detector array of 1024 pixels², 15µm each

Table 2 : Main top level requirements for collimating slicer study.

| Definition | Value | Unit |
|---|---|---|
| Telescope pupil diameter | 3.5 | m |
| Telescope F/D # | 3.93 | |
| FoV dimensions along X spatial direction | 0.20 | arcmin |
| FoV dimensions along Y spectral direction | 0.07 | arcmin |
| FoV angular resolution along X direction | 0.20 | arcsec |
| FoV angular resolution along Y direction | 0.20 | arcsec |
| Minimal wavelength | 350 | nm |
| Maximal wavelength | 750 | nm |
| Spectral resolution at reference wavelength | 200 | |
| Individual slice thickness | 0.5 | mm |
| Collimating Lens focal length | 100 | mm |
| Array along y-axis | 2 | |
| Array along x-axis | 10 | |
| Pixel size (on CCD) | 15 | µm |
| Pixel number along X spatial direction | 1024 | |
| Pixel number along Y spectral direction | 1024 | |
| Pixel number per sample along X direction | 1 | |
| Pixel number per sample along Y direction | 2 | |

Table 3 : Main subsystems requirements of the collimating slicer system.

| Definition | Value | Unit |
|---|---|---|
| **FORE-OPTIC CHARACTERISTICS** | | |
| Enlarger magnification along X&Y direction | 19x38 | |
| f#PreOptic | 74x148 | |
| **IMAGE SLICER CHARACTERISTICS** | | |
| Slice number | 20 | |
| Slicer mirror focal length | NA | |
| Individual slice dimensions | 15x0.5 | mm |
| Slice aspect ratio | 30 | |
| Slicers maximal half-angle around X direction | 6.5 | ° |
| Slicers maximal half-angle around Y direction | 2.25 | ° |
| Distance from slicer stack to collimating lens | 100 | mm |
| Distance from collimating lens to diffraction gr | 100 | mm |
| Intermediate Pupil plane diameter | 1.36x0.68 | mm |
| **SPECTROGRAPH CHARACTERISTICS** | | |
| Spectrometer magnification | 0.06 | |
| Collimator focal length | 100 | mm |
| Angle of Diffraction | 26.6 | ° |
| Groove | 1626 | Lines/mm |
| Resolving Power @λmin | 127 | |
| Resolving Power @λmax | 273 | |
| Camera focal length | 6 | mm |
| Camera F/D # | 4.4x8.8 | |
| Individual Spectra Size onto CCD in pixels | 0.9 x 4.5 | mm |

These main top level requirements are summarized in Table 2 and are considered as the starting point of the study. The desired system can be simply divided into fore-optics, image slicer and spectrograph subsystems whose major preliminary subsystem requirements are given in Table 3. This system analysis is essentially driven by their geometrical characteristics with the main goal of keeping in mind the system compactness and reducing the number and size of optical elements. Nevertheless as expected, a few arrangements had to be made:
- In order to match a slice thickness of 0.5mm and a spatial sampling of 0.2arcsec², the fore-optics has a magnification by ~19 x 38. Attaining a full numerical aperture F/D=74x148 at the image slicer entrance, the depth of focus is quite large and allows large tilts onto the slicer stack.

- Due to the FoV dimension and sampling along the Y spectral direction, the number of slices stacked together is 20. Each individual slice is 0.5 mm thick and 15 mm long giving a slice aspect ratio of 30 which is easily achievable for suppliers familiarized with optical contacting techniques. Each slice has different orientations (tilts) i.e. its surface centre is decentred along the Y' orthogonal axes following the arrangement in 2 rows of 10 images onto the detector. The maximal tilt is ±2.25° around the Y direction and the whole stack is tilted by +5° around the X direction to keep off the incoming beam. All slices are flat.
- The collimating lens is located at 100 mm from the slicer stack. Its focal length is fixed to $f_{Lens}$ = 100 mm in order to send the entrance FoV to infinity at the diffraction grating level.
- The diffraction grating has a diffraction angle of 26.6° with 1626 lines/mm to achieve the spectral resolution of 300. It is located at D' =100mm from the collimating lens due to flat slices. Each intermediate pupil measures 1.36 mm x 0.68 mm arranged along 2 rows of 10 intermediate pupil images.
- The microlens camera is also composed of 2 rows of 10 lenses. Each microlens works at F/#= 4.4 / 8.8 with a focal length of 6 mm. Its location should be as close as possible to the diffraction grating to minimize its size.
- The standard detector array has 1024 x 1024 pixels, 15µm each. It is located at the focal plane of the microlens camera. Each individual spectrum measures 0.9 x 4.5mm. All spectra are arranged in 2 rows of 10 spectra. The gap between two adjacent spectra is a few pixels.

The preliminary subsystem requirements in Table 3 are now sufficient to start optical designing of the collimating slicer with the help of Zemax ray tracing software. This is the scope of the next section.

### 4.3 Optical Design using Zemax

#### 4.3.1 Paraxial Lens Design

To demonstrate the feasibility of the concept, the previous optical design has been modeled with the Zemax software. This first approach uses paraxial lenses. It is composed of the fore-optics, slicer stack, collimating lens, diffraction grating, microlens camera and detector (see Figure 5).

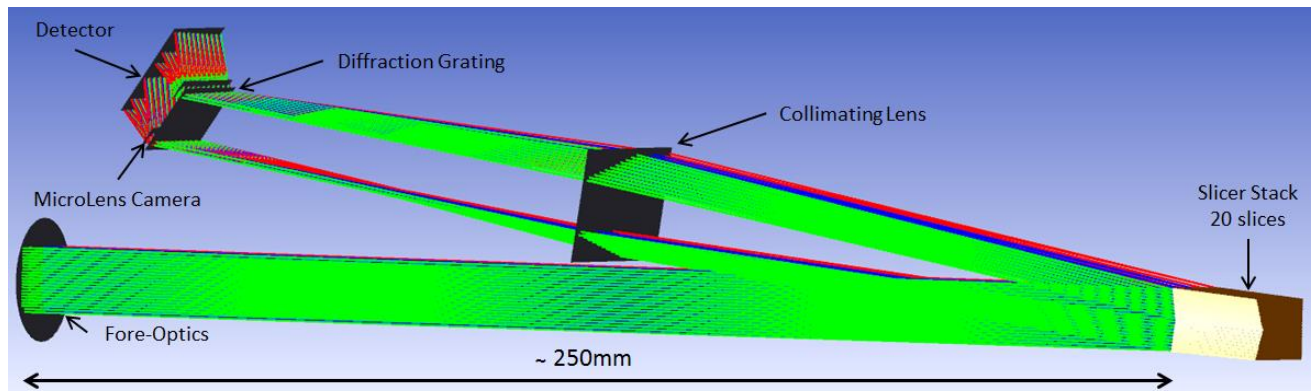

Figure 5 : Zemax model of the Collimating Slicer. For clearer view, the telescope is not represented on this figure.

The slicer stack is presented in Figure 6. The number of slices stacked together is 20. Each individual slice is 0.5 mm thick and 15 mm long. All slices have different orientations (tilts) in order to match the images arrangement onto the detector. The maximal tilt is ±2.25° around the Y direction. This tilt separates the intermediate pupils and fixes the final gap between the spectra onto the detector. It should be optimized. Tilts of +6.5° and +3.5° around the X direction are set to keep off the incident beam coming from the fore-optics and to separate each row of spectra onto the detector (Figure 5).

The focal length of the collimating lens is fixed to 100 mm. Its location is 100 mm away from the slicer stack in order to send the entrance FoV to infinity at the diffraction grating level. The footprint of the beams onto the collimating lens is described in Figure 7. For each slice, the intermediate pupil is shown as 3 elliptical sub-pupils corresponding to 3 fields on the slice (-7.5, 0 and +7.5 mm). The distance between collimating lens and diffraction grating were adjusted to have a common pupil onto the diffraction grating for each slice (Figure 8).

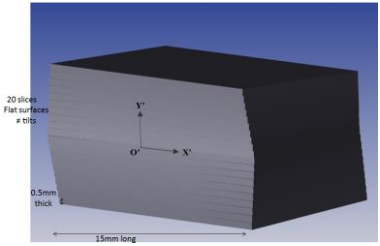

Figure 6 : Schematic view of the slicer stack

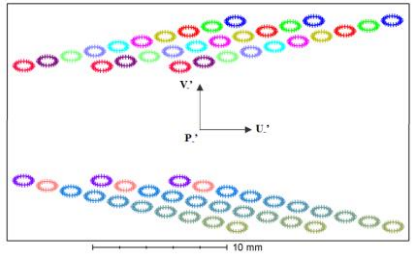

Figure 7 : Footprint onto Collimating Lens (one color per slice, 3 fields per slice)

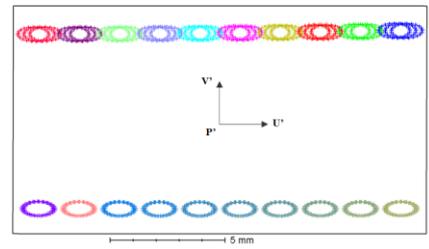

Figure 8 : Footprint onto each diffraction grating (one color per slice, 3 fields per slice)

The diffraction grating has a diffraction angle of 26.6° with 1626 lines/mm to achieve the required spectral resolution. It is located at ~75 mm from the collimating lens. Each intermediate pupil measures 1.36 mm x 0.68 mm arranged along 2 rows of 10 intermediate pupil images. Due to its diffraction angle of 26.6°, the locations of the 20 intermediate pupils are not at the same focus along the optical axis. The diffraction grating is centered onto the two rows of intermediate pupils (Figure 8).

The camera is composed of 2 rows of 10 microlenses each. Each microlens works at F/#= 4.4 / 8.8 with a focal length of 6 mm. Its location should be close as possible to the diffraction grating to minimize its size. In the next phase, it could be glued directly onto the diffraction grating. Each microlens of the camera has the same dimension (1.8 x 3 mm).

The standard detector array has 1024 x 1024 pixels, 15µm each. It is located at the focal plane of the microlens camera. Each individual spectrum measures 0.9 x 4.5mm (Figure 11). All spectra are arranged in 2 rows of 10 spectra. The gap between two adjacent spectra is ~30 pixels. The arrangement of spectra onto detector could be optimized in a next phase. The maximal spectral resolution achieved here is 300.

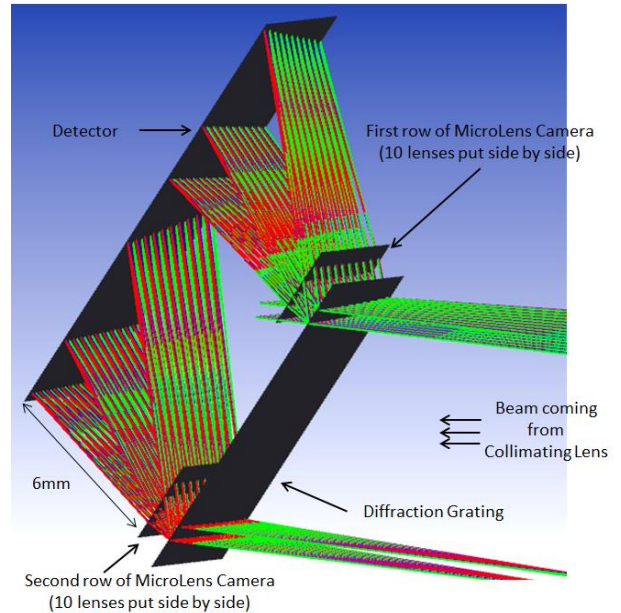

Figure 9 : Zoom view from the diffraction grating to the detector.

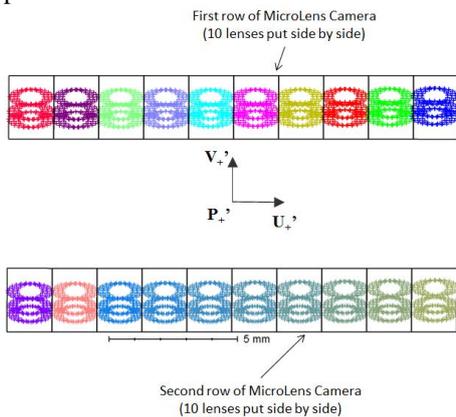

Figure 10 : Footprint onto the 2 rows of the microlens camera.

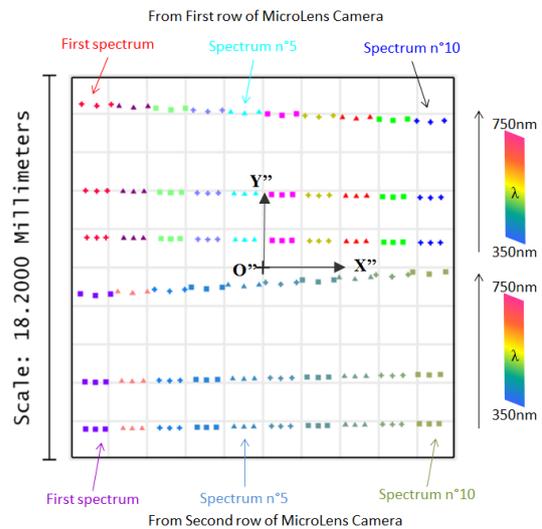

Figure 11 : Spectra onto the detector.

### 4.3.2 Bulk Optics Design

The previous paraxial lens design was used to define a preliminary optical layout with glass components. Here we essentially tried to minimize the deviation angle between the incident beam on the diffraction grating and the diffracted ray at the central wavelength of 550 nm, in order to minimize the geometrical aberrations of the micro lens camera. The resulting preliminary design is depicted in Figure 12 (to be compared with Figure 5). It is not yet finalized and does not fulfill all the technical specifications of section 4.2, especially spectral resolution and image quality requirements. However it already gives some precise ideas for what concerns overall dimensions, encumbrance, and the number and types of employed optical components. Its main features are the following:

- In order to decrease the deviation angle below 15 degrees while preserving the spectral resolution; the diffraction grating (1.48 lines/µm groove frequency) is associated with a 40.6 degrees SF4 prism, that can be cut into horizontal slices along the U'-axis in order to keep reasonable weight and dimensions (see the upper right vignette in Figure 12).
- A KZFSN4/FK51 achromatic doublet is used as the collimating lens located between the image slicer and diffractive elements.
- All the Fused Silica micro lenses used in a single row (along the U'-axis) of the micro lens camera are identical. However their geometrical parameters (curvature radii and thicknesses) were made variable from one row to the other.

Currently, the achieved image quality is insufficient, with a RMS spot radius on the detector array around 3-4 pixels depending on wavelength and field positions. However it can be improved in a few different ways:

- Optimize the choice of refractive materials and related parameters (including the substrate of the diffraction grating),
- Perform a coupled optimization with the fore-optics lens located upstream the image slicer,
- Eventually, reduce the required spectral range or shift it towards the red in order to benefit of more usable refractive materials.

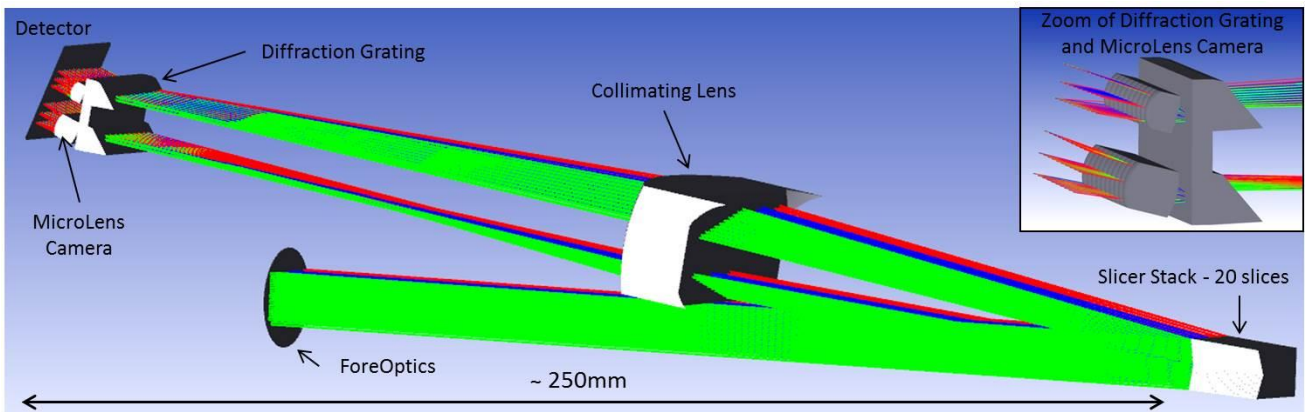

Figure 12 : Zemax model of the Collimating Slicer. For clearer view, the telescope is not represented on this figure.

## 5. CONCLUSION AND FUTURE DEVELOPMENT

In this paper, the state of art of Integral Field Spectroscopy in past and future astronomical instrumentations using image slicers has been introduced, summarizing different optical designs for image slicers. A new proposal for IFS called "Collimating Slicer" was analyzed. The idea is to combine the image slicer with the collimator of the spectrograph, thus mixing the "spatial" and "spectral" units. The classical combination of slicer, pupil and slit elements and spectrograph collimator is replaced with a new one composed of a slicer and spectrograph collimator only.

After testing few configurations, this collimating slicer looks very promising for low resolution spectrographs. Its main advantage is its compactness with 300 mm long typically. The main drawback is its low spectral resolution achieved due to small size of the re-imaged pupil onto the diffraction grating. This limitation could be alleviated by introducing a beam expander between the collimating lens and the diffraction grating. Moreover, a compromise could be found between the FoV and the spectral resolution as defined by the scientific case. The next steps will be to replace paraxial lenses by real optical components and start the realization of a prototype.

The collimating slicer is a compact module that can be coupled to other functionalities into future instrumentations for the next generation of telescopes implementing IFS.

## ACKNOWLEDGMENT

This work was supported by Rémi Boivineau during his engineer training course from ENSSAT, France. We acknowledge the financial support of the Labex Lyon Institut of Origin (LIO), France.

## REFERENCES

[1] Content et al., "A new design for Integral Field Spectroscopy with 8-m Telescopes," Proc. SPIE 2871, 1295 (1997)
[2] Thatte et al., "The E-ELT first light spectrograph HARMONI: capabilities and modes," Proc. SPIE 9908, 71 (2016)
[3] Lopez et al, "FRIDA, the diffraction limited NIR imager and IFS for the Gran Telescopio Canarias: status report," Proc. SPIE 9147, 1 (2014)
[4] Bowen, I.S., "The image slicer, a device for reducing loss of light at slit of stellar spectrograph," Ap.J., 88, 113 (1938)
[5] Pierce, A.K., "Construction of a Bowen Image Slicer," PASP 77, 216 (1965)
[6] Simmons et al., "Modified Bowen-Walraven Image Slicer," Proc. SPIE 331, 427 (1982)
[7] Richardson et al., "Image-slicers," Proc of the IAU Colloquium, 79, 469 (1984)
[8] Weitzel et al., "3D: The next generation near-infrared imaging spectrometer," A&AS 119, pp. 531–546 (1996)
[9] Laurent et al., "Designing, manufacturing and testing of an advanced image slicer prototype for the James Webb Space Telescope," Proc. SPIE, 5494, 196 (2004)
[10] F. Hénault, R. Bacon, R. Content, B. Lantz, F. Laurent, J.-P. Lemonnier, S. Morris, "Slicing the Universe at affordable cost: The Quest for the MUSE Image Slicer," Proceedings of the SPIE vol. 5249, p. 134-145 (2003).
[11] F. Laurent, F. Hénault, E. Renault, R. Bacon, J.-P. Dubois, "Design of an Integral Field Unit for MUSE, and results from prototyping," Publications of the Astronomical Society of the Pacific vol. 118, n° 849, p. 1564-1573 (2006).
[12] Laurent et al., "Innovative slicer design and manufacturing," Proc. SPIE, 6273, 71 (2006)
[13] Laurent F., et al., "MUSE integral field unit: test results on the first out of 24, " Proc. SPIE, 7739, 147 (2010)
[14] Calcines et al., "MuSICa: The multi-slit image slicer for the EST spectrograph," Journal of Astronomical Instrumentation, 2, 1 (2013)